\input harvmac

%%%%%%%%%%%%%%%%%%%%%  Rublenye bukvy   %%%%%%%%%%%%%%%%%%%%%%%%
\def\IB{\relax\hbox{$\inbar\kern-.3em{\rm B}$}}
\def\IC{\relax\hbox{$\inbar\kern-.3em{\rm C}$}}
\def\ID{\relax\hbox{$\inbar\kern-.3em{\rm D}$}}
\def\IE{\relax\hbox{$\inbar\kern-.3em{\rm E}$}}
\def\IF{\relax\hbox{$\inbar\kern-.3em{\rm F}$}}
\def\IG{\relax\hbox{$\inbar\kern-.3em{\rm G}$}}
\def\IGa{\relax\hbox{${\rm I}\kern-.18em\Gamma$}}
\def\IH{\relax{\rm I\kern-.18em H}}
\def\IK{\relax{\rm I\kern-.18em K}}
\def\IL{\relax{\rm I\kern-.18em L}}
\def\IP{\relax{\rm I\kern-.18em P}}
\def\IR{\relax{\rm I\kern-.18em R}}
\def\IZ{\relax\ifmmode\mathchoice
{\hbox{\cmss Z\kern-.4em Z}}{\hbox{\cmss Z\kern-.4em Z}}
{\lower.9pt\hbox{\cmsss Z\kern-.4em Z}}
{\lower1.2pt\hbox{\cmsss Z\kern-.4em Z}}\else{\cmss Z\kern-.4em
Z}\fi}

\def\II{\relax{\rm I\kern-.18em I}}

%%%%%%%%%%%%%%%%%%%%% Calligraphic letters  %%%%%%%%%%%%%%%%%%%%%

\def\CD {{\cal D}}

\def\CF {{\cal F}}

\def\CM {{\cal M}}
\def\CN {{\cal N}}

\def\CS {{\cal S}}

\def\CX {{\cal X}}

%%%%%%%%%%%%%%%%%%%%%%%%%% Derivatives  %%%%%%%%%%%%%%%%%%%%%%%%

\def\p{\partial}

%%%%%%%%%%%%%%%%%%%% letters with bar %%%%%%%%%%%%%%%%%%%%%%%%%%

%%%%%%%%%%%%%%%%%%%%%%%%%%% Math symbols %%%%%%%%%%%%%%%%%%%%%%%

\def\p{\partial}

\def\inbar{\,\vrule height1.5ex width.4pt depth0pt}
\font\cmss=cmss10 \font\cmsss=cmss10 at 7pt

%%%%%%%%%%%%%%%%%%%%%%%%%%%%%%%%%%%%%%%%%%%%%%%%%%%%%%%%%%%%%%%%%
\def\a{\alpha}

\def\G{\Gamma}
\def\b{\beta}

\def\e{\epsilon}

\def\f{\varphi}
\def\p{\partial}

\def\Z{\relax\ifmmode\mathchoice
{\hbox{\cmss Z\kern-.4em Z}}{\hbox{\cmss Z\kern-.4em Z}}
{\lower.9pt\hbox{\cmsss Z\kern-.4em Z}}
{\lower1.2pt\hbox{\cmsss Z\kern-.4em Z}}\else{\cmss Z\kern-.4em Z}\fi}

\def\r{\relax{\rm I\kern-.18em R}}
\font\cmss=cmss10 \font\cmsss=cmss10 at 7pt

\def\CX{{\cal X}}
\def\CM{{\cal M}}

%--------+---------+---------+---------+---------+---------+---------+
%Steve's macros: these seem to work both in latex and harvmac.
%To produce a box for a Dalembertian (adapted from p. 320 of TeXbook):
\def\sqr#1#2{{\vcenter{\vbox{\hrule height.#2pt
         \hbox{\vrule width.#2pt height#1pt \kern#1pt
            \vrule width.#2pt}
         \hrule height.#2pt}}}}

%Extra space here looks nicer in main math text mode.
%--------+---------+---------+---------+---------+---------+---------+
%%%REFERENCES
%%
%%% Kontsevich
\lref\enumer{M.~Kontsevich, ``Enumeration  of rational curves via torus actions'', In: {\it 
The moduli 
space of curves}, ed. by R.~Dijkgraaf, C.~Faber, G.~van der Geer,
Progress in Math. vol. 129, Birkh\"auser, 1995, 335-368}
%%% Manin
\lref\manineu{Yu. Manin, ``Generating functions in algebraic geometry and sums over trees'',
alg-geom/9407005. In: {\it 
The moduli 
space of curves}, ed. by R.~Dijkgraaf, C.~Faber, G.~van der Geer,
Progress in Math. vol. 129, Birkh\"auser, 1995, 401-418}
\lref\manineum{Yu. Manin, ``Stable maps of genus zero to flag spaces'', math.AG/9801005}
%%% Penner model
\lref\penner{J.~Harer, D.~Zagier, ``The Euler characteristics of the moduli space of curves'', 
Inv. Math. {\bf 85} (1986), 457-485 \semi
R.~C.~Penner, ``Perturbative series and the moduli space of Riemann surfaces'',  J.~Diff.~Geom. {\bf 27} (1988), 35-53 \semi
J.~Distler, C.~Vafa, ``The Penner model and $d=1$ string theory'', in, Proc. of the Carese Workshop
on Random Surfaces, Quantum Gravity and Strings, May 1990\semi
E.~Getzler, M.~Kapranov, ``Modular Operads'', dg-ga/9408003}
%%%%
\lref\vafabl{C.~Vafa,
``Black Holes and Calabi-Yau Threefolds''
hep-th/9711067,  Adv. Theor. Math. Phys. {\bf 2} (1998) 207-218}
\lref\CMR{S.~Cordes, G.~Moore, and S.~Ramgoolam,
`` Lectures on 2D Yang Mills theory, Equivariant
Cohomology, and Topological String Theory,'', 
hep-th/9411210, or see http://xxx.lanl.gov/lh94}
\lref\dnld{S. Donaldson, ``Anti self-dual Yang-Mills
connections over complex  algebraic surfaces and stable
vector bundles,'' Proc. Lond. Math. Soc,
{\bf 50} (1985)1}
\lref\vkm{A.~Klemm, P.~Mayr, C.~Vafa, ``BPS States of Exceptional Non-Critical Strings'', hep-th/9607139}
\lref\albnik{A.~Lawrence, N.~Nekrasov, ``Instanton sums and five dimensional gauge theories'', Nucl. Phys. {\bf B} 513 (1998) 239-265, hep-th/9706025}
\lref\gopvaf{M.~Marino, G.~Moore, ``Counting higher genus curves in a Calabi-Yau manifold'', hep-th/9808131\semi
R.~Gopakumar, C.~Vafa, ``$M$-theory and topological strings-I'', hep-th/9809187}
\lref\witpha{E.~Witten, ``Phase Transitions in   $M$-theory and $F$-theory'', hep-th/9603150}
\lref\sadvaf{V.~Sadov, C.~Vafa, unpublished}
\lref\local{S.~Katz, A.~Klemm, C.~Vafa, ``Geometrical Engeneering of Quantum Field Theories'',
hep-th/9609239}
\lref\nikbarak{B.~Kol, N.~Nekrasov, in progress}
%%%%%%%%%%%%%%%%% Witten %%%%%%%%%%%%%%%%%%%%%%%%%%%%%%%%
\lref\witdyn{E. Witten, ``String theory dynamics
in various dimensions,''
Nucl. Phys. {\bf B} 443 (1995) 85-126}
\lref\WitDonagi{R.~ Donagi, E.~ Witten,
``Supersymmetric Yang-Mills Theory and
Integrable Systems'', hep-th/9510101, Nucl. Phys.{\bf B}460 (1996)
299-334}
\lref\witbound{E.~Witten, ``Bound States Of Strings And $p$-Branes'',
hep-th/9510135
Nucl. Phys. {\bf B}460 (1996) 335-350}
\lref\witconst{E.~Witten, ``Constraints on supersymmetry breaking'',
Nucl. Phys. {\bf B}202 (1982) 253}

\lref\nakheis{H.~Nakajima,
``Lectures on Hilbert schemes of points on surfaces'', H.~Nakajima's
homepage}
\lref\vw{C.~Vafa, E.~Witten, ``A strong coupling test of
$S$-duality'', hep-th/9408074;
Nucl. Phys. {\bf B} 431 (1994) 3-77}
\lref\dvafa{C.~Vafa, ``Instantons on D-branes'', hep-th/9512078,
Nucl. Phys. B463 (1996) 435-442}
\lref\atbott{M.~Atiyah, R.~Bott, ``The Moment Map And
Equivariant Cohomology'', Topology {\bf 23} (1984) 1-28}
\lref\atbotti{M.~Atiyah, R.~Bott, ``The Yang-Mills Equations Over
Riemann Surfaces'', Phil. Trans. R.Soc. London A {\bf 308}, 523-615
(1982)}
\lref\issues{A.~Losev, N.~Nekrasov, S.~Shatashvili, 
``Issues in Topological Gauge Theory'', 
hep-th/9711108,  Nucl.Phys. {\bf B}534 (1998) 549-611}
\lref\rahulp{M.~Kontsevich, talk at the Int. Congress on Math. Physics,
Paris, 1994\semi
 T. Graber, R. Pandharipande, ``Localization of virtual classes'',
alg-geom/9708001}
%%%%%%%%%%%%%%%%%%%%%%%%%%%%%%%%%%%%%%%%%%%%%%%
\Title{ \vbox{\baselineskip12pt\hbox{hep-th/9810168}\hbox{HUTP- 98/A074}
\hbox{ITEP-TH-58/98}}}
{\vbox{ \hbox{In the Woods of M-theory}
}}
%%%%%%%%%%%%%%%%%%%%%%%%%%%%%%%%%%%%%%%%%%%%%%%%%
\centerline{Nikita Nekrasov}
\footnote{}{nikita@curie.harvard.edu}

\vskip .4in

\centerline{\it Institute
 of Theoretical and Experimental
Physics, 117259, Moscow, Russia}

\centerline{\it  Lyman Laboratory of Physics,
Harvard University, Cambridge, MA 02138, USA}

\vskip .4in

We study BPS states  which arise in compactifications
of $M$-theory on Calabi-Yau manifolds. In particular, we are interested
in the spectrum of the particles obtained by wrapping $M2$-brane
on a two-cycle in the CY manifold $X$. We compute  the Euler characteristics of the moduli space
of genus zero curves which land in a holomorphic four-cycle $S \subset X$. We use M.~Kontsevich's method which reduces the problem to  summing over trees and observe
the discrepancy with the predictions of local mirror symmetry. 
We then turn this discrepancy into a supporting evidence in 
favor of  existence of extra moduli of $M2$-branes which consists
of the choice of a flat $U(1)$ connection
recently suggested by  C.~Vafa and  partially confirm this by counting of the arbitrary genus curves of bi-degree $(2,n)$ in $\IP^1 \times \IP^1$ (this part has been done together with Barak Kol). We also make a conjecture concerning the counting of higher genus
curves  using second quantized Penner model  and discuss possible applications to the string theory of two-dimensional $QCD$.

\bigskip

\Date{10/98}

\newsec{Introduction }

The study of $M$-theory \witdyn\  is a fascinating problem. One of the hints
towards its existence is the hidden five dimensional structure
in the geometry of four dimensional $\CN=2$ gauge theories.
For example, if the gauge theory is realized as Type $IIA$ string
theory compactified on a (perhaps singular) Calabi-Yau manifold
$X$ then its vector multiplet structure is encoded in the prepotential:
\eqn\prepi{\CF = \sum_{\vec n} N_{\vec n} {\rm Li}_{3}(e^{-\vec n \cdot \vec t})}
where $\vec t$ are the special coordinates on the Kahler moduli space of $X$,
$\vec n$ runs through  a set of integral points in the positive Kahler
cone and $N_{\vec n}$ is the ``number''
of the  rational curves representing the homology
class $\vec n$. The word ``number'' is in quote marks since
it may be rational due to singularities of the moduli space of curves. 

The promised five dimensional structure is in the fact that when translated
to the effective coupling the expression \prepi\ can be written as follows:
\eqn\effcpl{\tau_{ij} = {{\p \CF}\over{\p t_{i} \p t_{j}}} =
\sum_{\vec n} n_{i} n_{j} N_{\vec n} \sum_{k \in \IZ} {\rm log}
\left( \vec t \cdot \vec n + 2\pi  k \right) + {\rm log} \Lambda }
In this representation we replaced
$Li_{1}(e^{-a}) \equiv - {\rm log}(1 - e^{-a})$ by
$$
{{a}\over{2}} - {\rm log} \left( 2 {\rm sinh} {{a\over{2}}} \right) =
{{a}\over{2}} - \sum_{k \in \IZ} {\rm log} \left( 2\pi n  + a \right) + {\rm log} \Lambda
$$
where $\Lambda$ is a divergent constant. This representation is most naturally interpreted
as a sum over all Kaluza-Klein charged states \albnik.

The paper is organized as follows. In the section $2$ we remind the setup of local 
mirror symmetry and describe the non-compact Calabi-Yau manifolds which we
are going to study. In the section $3$ we proceed with counting of curves in genus zero. 
The section $4$ (speculative at the moment) contains a comparison to the predictions of mirror
symmetry and partial resoltuion of the contradiction which appears in the course of such comparison. The section $5$ contains concluding remarks and speculations about counting
of higher genus curves, applications to the string theory of $QCD_{2}$, and other remarks.

\newsec{The setup}

\subsec{The local geometry}

Consider 
the Calabi-Yau manifoild $X$ which contains a surface $S$ with $b_{2}^{+}=1$.
The normal bundle to $S$ is necessarily isomorphic to its canonical
bundle $K_{S}$ and the latter has no sections, since 
$2 h^{2,0}(S) = b_{2}^{+} - 1 = 0$. In computing the prepotential of the four dimensional
theory obtained by compactification of $\II$A string on $X$ one encounters a
problem of counting the holomorphic curves in $X$ which land in $S$.
When lifted to $M$-theory the same question arises in the study
of BPS particles obtained by wrapping the $M2$-brane around two-cycles
$\beta$ in $S$.

For the generic complex structure on $X$ there are no holomorphic
surfaces $S$ embedded into $X$. Also, the holomorphic
curves in $X$ are generically isolated. What it means is that the problem of
counting curves in $X$ and the problem of counting curves in $S$ are different.
Nevertheless, it is clear that given the complex structure on $X$ which is such that
$S$ is a holomorphic submanifold in $X$ one encounters a {\it moduli}
space $\overline\CM = \amalg \overline\CM_{\beta}$ of curves landing in $S$. It turns out that on top of this
moduli space one also encounters the {\it obstruction bundle}
$\CF = \amalg \CF_{\beta}$. The meaning of this bundle is roughly the following. 
Take a holomorphic curve in $S$ and start varying the complex
structure of $X$. The generic curve will cease to remain holomorphic.
Only some special points in $\CM_{\beta}$  correspond to the curves which 
remain holomorphic. These points
can be identified with the zeroes of some section of $\CF_{\beta}$. The number of
zeroes counted with appropriate multiplicity is nothing but
the Euler number of the bundle $\CF_{\beta}$.
Hence the prepotential \prepi\ receives a contribution which is
proportional to 
\eqn\elr{N_{\beta} = \int_{\overline\CM_{\beta}} {\rm Eu} {\CF_{\beta}}}
It is important that ${\rm rk}\CF_{\beta} = {\rm dim}\overline\CM_{\beta}$ otherwise $N_{\beta} = 0$. 

\subsec{The strategy}

We are going to study the toric  $S$. The variety $S$ is acted upon by the two dimensional torus $T^{2}$. The quotient $S/ T^{2}$ is a convex
polygon $\Delta$ due to a theorem by 
M.~Atiyah and has the following meaning.  Its vertices $f$  correspond to the fixed points
of the torus action while the edges connecting two vertices $f_{1}, f_{2}$ correspond to the
degenerate orbits. The union of all degenerate orbits which sit on top of
the given edge is isomorphic to a two-sphere. Let $\vec d_{f_{1}, f_{2}}$ be the 
homology class of this sphere. We denote by $t_{f_{1}, f_{2}} = \vec t \cdot
\vec d_{f_{1}, f_{2}}$. 

The torus $\bf T = \IC^{*} \times \IC^{*}$ which is a complexification of $T^{2}$
acts on $S$. It  also acts on the moduli space of stable maps of the curves into $S$.
The latter is a disjoint union of the spaces $\overline\CM_{g, \beta}$
where $g$ denotes the genus  of a generic
curve $C$ and $\beta$ the degree of the map: $\beta  = [C] \in H_{2}(S)$.

The fixed points of the action of $\bf T$ on $\overline\CM_{g,k, \beta}$ were described
by M.~Kontsevich in \enumer. We first describe them in words and then proceed
to enumerating them. Roughly speaking the only possibility
for the stable curve to be fixed under the $\bf T$ action is to be a union of
genus zero curves which are mapped to the spheres connecting the fixed points in $S$ and 
the rest of the components mapped to the fixed points. In other words the normalization
$\widetilde C$ of $C$ splits as follows:
\eqn\dscr{{\widetilde C} = \amalg_{\alpha} C_{\alpha}}
where for each $C_{\alpha}$ there are  two options. If $C_{\a}$  has genus higher then zero or it has
more then two special (marked or singular) points, then it is mapped to one of the
fixed points $f \in S$. If $C_{\alpha}$ is of genus zero and has exactly two 
special points then it is mapped to the sphere connectiong $f_{1}, f_{2}$
where $f_{1}$ and $f_{2}$ are the fixed points in $S$ which are at the same time
the images of the two special points in $C_{\alpha}$. The component of $\overline\CM_{g, \beta}^{\bf T}$ which contains this curve is isomorphic to $\prod_{v} \overline\CM_{g_{v}, n_{v}}$. Here the product is taken over all components which are mapped to points. 
The paper \enumer\ explicitly describes all components of the fixed set. They are enumerated by the
equivalence classes of the connected graphs $\Gamma$ ($E_{\G}$, $V_{\G}$ denote respectively
the set of edges and the set of vertices of $\G$) with the following data. 

\item{1.}To each vertex $v \in V_{\G}$
 a vertex in $\Delta$ is assigned : $v \mapsto f_{v}$. 
\item{2.}To each edge $e$ an integer $d_{e} > 0$ is assigned.
\item{3.}If an edge connects two vertices $v_{1}, v_{2}$ then $f_{v_{1}} \neq f_{v_{2}}$.
\item{4.} $\sum_{e \in E_{\G}} d_{e} \cdot \vec d_{f_{v_{1}}, f_{v_{2}}}=  \vec d$
\item{5.} To each vertex $v$ a set $S_{v} \subset \{1, \ldots k \}$ of the marked points 
which belong to the component $C_{v}$ of a curve is assigned. $\{ 1, \ldots, k \} = \amalg_{v \in V_{\G}} S_{v}$
\item{6.} To each vertex a number $g_{v}$ (an internal genus) is assigned. The stability implies that $ 2 - 2g_{v} - n_{v} < 0$.

\noindent
The total genus of the curve $C$ is computed as
\eqn\totalg{g =1 + \# E_{\G} + \sum_{v \in V_{\G}} ( g_{v}  - 1)  }

\newsec{Counting in  genus zero}

\subsec{Preliminary construction without marked points}

If we restrict ourselves to the genus zero maps $g=0$ then the graphs are necessarily
trees and our computation reduces to the sum over all trees of the certain weights. 
The vertex $v$ of valency $n_{v}$ contributes a factor $\chi ( \overline\CM_{0, n_{v}} )$.
Let us introduce a notation: $\chi (\overline\CM_{0,1} ) = \chi (\overline\CM_{0,2} ) = 1$.
Then the sum of all contributions of the fixed components can be written as:
\eqn\expr{\sum_{\Gamma, n, \vec d} {1\over{\# {\rm Aut}({\Gamma})}} \prod_{v \in V_{\Gamma} } 
\chi ( \overline\CM_{0, n_{v}} ) \prod_{e \in E_{\Gamma}} e^{- \vec t \cdot \vec d_{e}} }  

This is nothing else but the critical value of the action functional:
\eqn\actn{S( \phi) = -\half \sum_{d, \langle i j\rangle } e^{ t_{ij} d} \phi_{ij,  d} \phi_{ji, d}  + \sum_{k \geq 1}  {1\over{k!}} \chi ( \overline\CM_{0,k} )
\sum_{i, \{ ( j_{1}, d_{1}) \ldots (j_{k} , d_{k}) \} } \phi_{ij_{1},  d_{1} }
\ldots \phi_{ij_{k}, d_{k}} }
where $i,j$ label the vertices of the polygon (or the fixed points of the torus action on $S$),
the ``field'' $\phi_{ij, d}$ is non-zero only iff $i$ and $j$ are connected by an edge, 
$d \in \bf N$. 
Introduce the following notations:
\eqn\nts{\eqalign{\phi_{i} = & \sum_{j,  d} \phi_{ij, d} \cr
\chi(\f) =  & \sum_{k \geq 1} \chi ( \overline\CM_{0,k} )
{{\f^{k}}\over{k!}}  =  \f + {{\f^{2}}\over{2}} + {{\f^{3}}\over{6}} + \ldots \cr }}
Then the action \actn\ can be written much more simply as:
\eqn\actni{S = - \half \sum_{d, \langle i j\rangle } e^{ t_{ij} d} \phi_{ij, d} \phi_{ji, d} + \sum_{i} \chi ( \phi_{i}) } 
Its critical points are the solutions to the following system of equations:
\eqn\sstm{\eqalign{& \phi_{ji, d} =  e^{- t_{ij}  d} \xi_{i} \cr
&\xi_{i} =   \chi^{\prime} ( \phi_{i}) \cr
&\phi_{i} =  \sum_{j \neq i, \langle i j \rangle}  {1\over{e^{t_{ji}} -  1}} \xi_{j} 
\cr}}
In \manineu\ the following representation for $\chi(\f)$ can be found:
\eqn\mdl{\eqalign{\chi(\f) = & \f +  {\rm Crit}_{\xi}  \left( {\CS}(\xi) + \xi \f \right)\cr
{\CS} (\xi) = &  {\half} (1 + \xi)^{2} {\rm log} (1 + \xi ) - {{\xi}\over{2}} - {{5{\xi}^{2}}\over{4}} \cr}}
Hence we can replace the action \actn\ by the new action which depends on extra variables $\xi_{i}$ without affecting its critical value:
\eqn\bigac{
S( \phi) \to   -\half \sum_{d, \langle i j\rangle } e^{ t_{ij} d} \phi_{ij,  d} \phi_{ji, d} +
\sum_{i} \left( 1 + \xi_{i} \right) \phi_{i} + {\CS} ({\xi}_{i})}
Now, since this action depends quadratically on $\phi_{ij,d}$ they can be eliminated, leaving us with the action which depends (non-linearly) only on a finite number of variables:
\eqn\seff{{\Theta} ( \xi) = {\half} \sum_{\langle i j \rangle} {{( 1 + {\xi}_{i} )( 1 + {\xi}_{j})}\over{e^{t_{ij}} - 1}} + \sum_{i} {\CS} ({\xi}_{i})}
Notice that this is in a way the opposite to what was done in \enumer\ in a similar situation.

\subsec{Example: the projective space}
Consider $S = \IP^{n}$. The polygon $\Delta$ is the simplex with $n+1$ vertices $f_i$, labelled
by $i=0, \ldots, n$. All $t_{ij}$ are equal to each other. Introduce the notation:
$z = e^{- t_{ij}}, Q = {z\over{1 - z}}$. Then the formal minimum of the action \seff\ is obtained by setting all $\xi_{i}$
to be equal to each other:
\eqn\seffp{{\Theta} ({\xi} ) =Q {{n ( n+1)}\over{2}} ( 1 + {\xi} )^{2} + (n+1) \CS ({\xi})}
The critical point of \seffp\ obeys:
\eqn\crptn{n Q ( 1 + {\xi} ) + ( 1 + {\xi}) {\rm log} ( 1  + {\xi} )  - 2{\xi} = 0}
This equation is to be compared to the $t=0$ case in \manineum\ (The equation (2.11) there
yields the same equation for the critical point as \crptn. Moreover,
the critical value agree with the equation (2.1) of \manineum ). 

\subsec{Maps with marked points}

Suppose we are interested in counting the stable maps with $k$ marked
points on the worldsheet. The stability condition involves the marked points,
so the answer will be slightly different. For example, a curve without marked points has no stable maps of degree zero, while the curve with three marked points has.
It is easy to incorporate marked points, though. 
In the language of graphs we used so far the marked points can be thought as being the extra vertices, such that the edges which connect them to the rest of the graph carry no index $d$. 
This can be incorporated by the shift  $\phi_{i} \to \phi_{i} + t$ 
in  the action \actni\ together with the shift of the whole 
action by $\sum_{i}  - t - {{t^{2}}\over{2}}$ in order to remove the unstable configurations with 
only two or one marked point on the component which is sent to a point in the target space:
\eqn\mdfactn{S = - \half \sum_{d, \langle i j\rangle } e^{ t_{ij} d} \phi_{ij, d} \phi_{ji, d} + \sum_{i} \left( \chi ( \phi_{i} + t )  - t - {{t^{2}}\over{2}} \right)}
The effective action becomes:
\eqn\seffi{{\Theta} ( \xi) = {\half} \sum_{\langle i j \rangle} {{( 1 + {\xi}_{i} )( 1 + {\xi}_{j})}\over{e^{t_{ij}} - 1}} + \sum_{i} \left( {\CS} ({\xi}_{i}) + t\xi_{i} - {{t^{2}}\over{2}}  \right)}
The critical point now obeys:
\eqn\crptni{ ( 1 + {\xi}_{i}) {\rm log} ( 1  + {\xi}_{i} )  - 2{\xi}_{i} +  t = - \sum_{\langle ji \rangle}
{{( 1 + {\xi}_{j})}\over{e^{t_{ij}} - 1}} 
}
In the example of $S = \IP^{n}$ this reduces to 
\eqn\crpn{n Q ( 1 + {\xi} ) + ( 1 + {\xi}) {\rm log} ( 1  + {\xi} )  - 2{\xi}  +  t = 0}
which again agrees with  the equation (2.11) in \manineum.
It is also possible to express the critical value of the action as a polynomial in  $\xi_{i}$:
\eqn\crvl{{\rm Crit} {\Theta} = {1\over{4}} \sum_{i} \left( 1 - t^{2} - ( 1+ t - \xi_{i})^{2} \right)}
This is the main result so far:
\eqn\mnresult{\sum_{k, \vec d} \chi (\overline\CM_{0,k, \vec d}) {{t^{k}}\over{k!}} e^{- \vec t \cdot \vec d} = {1\over{4}} \sum_{i} \left( 1 - t^{2} - ( 1+ t - \xi_{i})^{2} \right)}
As a simple check, in the limit $t_{ij} \to \infty$ (degree zero maps) we get just the 
Euler characteristics of the moduli spaces of curves $\overline\CM_{0,k}$ times the euler
characteristics of the target $S$ (again in agreement with (2.11)  in \manineum\ ).

In the sequel we shall use the identity:
\eqn\frstde{{{d {\rm Crit}\Theta}\over{d t}} = \sum_{i} \left( \xi_i - t \right)}

\newsec{Comparison to  the results of mirror symmetry}

We compare the results of the previous section to the predictions of the local mirror symmetry in the case: $S = \IP^2$. The formula \crvl\ implies:
\eqn\ptwoid{\widetilde\CF_{0} (z) := \sum_{d} \chi (\overline\CM_{0,0, d}) z^{d} ={3\over{2}}  \left(\xi - {1\over{2}} \xi^2 \right)}
where $\xi$ solves the equation:
\eqn\ptwoe{f_{z}(\xi) =  {{2z}\over{1- z}}, \quad f_{z} (\xi) = 2\left( 1 - {z\over{1- z}}\right) {\xi} - ( 1 + {\xi}) {\rm log} ( 1  + {\xi} )   = 0}
It is easy to expand the solution to \ptwoe\ as a series in $z$: first write
\eqn\expa{\xi = \sum_{k=0}^{\infty} \left({{2z}\over{1- z}} \right)^{k}  a_{k}(z)}
where
\eqn\expaa{a_{k}(z) = {\rm Res} \, {\xi} {{d f_{z}(\xi)}\over{f_{z}^{k+1}(\xi)}}}
and then re-expand in $z$. 
Explicitly we get:
\eqn\ptwoexpl{\widetilde\CF_{0} (z) = 3z + 9z^2 + 31 z^3+ 121  z^4 + {{2623}\over{5}}  z^5 + {{37111}\over{15}} z^6 + \ldots}
We may also consider the generating function for 
the Euler characteristics of the moduli spaces of curves with three marked points. In this case we get:
\eqn\ptwomrk{\widetilde\CF_{3}(z) = 3 + 30 z + 276 z^2 + 2398 z^3 + 20174 z^4 + 166266 z^5 + {{20263976}\over{15}} z^6 + \ldots  }
On the other hand \vkm\ got the expression (in our normalization):
\eqn\excep{\CF (z) = 3z  + 9 z^2 + 30z^3 + 201z^4 + 1698 z^5 + 17100 z^6 + \ldots}
(we added multiple coverings to the numbers we got from the Table 6 for $d=0$ in \vkm). 
We see that the numbers \excep\ are slightly bigger then those in \ptwoexpl\ but are much 
smaller then those in \ptwomrk. 
All this implies (together with the fact that the Euler characteristics comes out fractional
for high degree maps, as the moduli spaces are really orbifolds) that there is a discrepancy and that the moduli space
of wrapped $M2$-brane is different from merely the moduli space of
holomorphic curves. Indeed, it was shown in \albnik\ (using the results of \witpha) that if $\CM$ is the space
of the collective coordinates of the soliton in $M$-theory which gives rise to a BPS
particle in five dimensions then its contribution to the four dimensional beta function and
correspondingly to the prepotential $\CF$ is precisely $\chi (\CM)$. 
The resolution to this puzzle was  suggested  by C.~Vafa \vafabl.
The claim is that one should think of the wrapped $M2$-brane as 
of the wrapped $D2$-brane.
The latter has extra moduli which come from the $U(1)$ gauge field on 
the brane world-volume.

Here we present a few arguments in favor of this conjecture. The mysterious as it seems such an addition is natural if we try to perform the duality 
transformation which maps the gauge field to the extra scalar describing the motion of
the brane in eleventh dimension more carefully. Since the duality transformation operates on the curvature of the gauge field: $F = \star d\phi_{D}$ it misses precisely the flat connections, i.e.
the gauge fields $A$ whose curvature vanishes: $F = dA = 0$. 

Since the Jacobian (the moduli space of flat connections) of the genus $g$ curve is birationaly
equivalent to the $g$'th symmetric power of the curve it is natural to try to parameterize
the moduli space of pairs $(C, L)$, where $C$ is a holomorphic curve and $L$ is a holomorphic
line bundle on it by first choosing $g$ points in $S$ and then considering the space of curves which pass through these points. The latter space is isomorphic to $\IP^{d}$
where $d$ is the dimension of the moduli space of rational curve in a given homology class $\b$:
$d = \langle c_{1}(S), \b \rangle - 1$. 

One evidence in favor of this  proposal comes from the study of the BPS states in the compactification of $M$-theory on K3 where one expects to see the spectrum of heterotic
string compactified on $T^3$. In this case the virtual dimension of the moduli space of rational 
curves is negative. The symmetric product of K3's can be resolved to a smooth
hyperkahler manifold, whose Euler characteristics is generated by the series:
$$
\sum_{g} q^{g} \chi \left( \widetilde{{\rm Sym}_{g} ({\rm K3})} \right)  = {1\over{\prod_{n} ( 1- q^{n})^{24}}}
$$
where one can recognize the right-moving oscillators of the heterotic string.  This example shows that the symmetric product of the manifold has to be resolved (this resolution is needed
for two purposes: first - in order to have a curve which passes through the coincident points one needs the tangent line to the curve at the point, and second - we mentioned above that
the Jacobian of the curve is not quite the symmetric product - they are related by  a 
birational transformation). Unfortunately at this point we are unable to state precisely
what is the resolution of the symmetric product which one has to consider. We present
one more example (which was discovered by  joint efforts with B.~Kol), namely the curves in 
$S = \IP^1 \times \IP^1$. The curves have bi-degree $(m,n)$ - the number of times
they wrap the first and the second $\IP^1$'s respectively. The dimension $d$ in this case is equal to $2(m+n)$. The genus of generic curve in the $(m,n)$ homology class is $g = (m-1)(n-1)$.  The symmetric products of $S$ (without resolution) have Euler characteristics
which is generated by the series
$$
\sum_{g} q^{g} \chi \left( {{\rm Sym}_{g} (S)} \right)  = {1\over{( 1- q)^{4}}}
$$
(i.e. $4$ bosonic oscillators instead of the $4$ right-moving chiral bosons as would have been the
case for the complete resolution). It turns out that the partial resolution
\eqn\part{{\CX} (q) = \sum_{g}  q^{g} \chi \left(\widehat{{\rm Sym}_{g} (S)} \right)  = {1\over{( 1- q)^{4}(1-q^2)}}
} does the job! More precisely, the series
\eqn\prepo{\eqalign{& \CF_{(2,*)} (z, w) =  z^{2} \sum_{n} w^{n} \left( 2(n+2)  \chi \left(\widehat{{\rm Sym}_{n-1} (S)} \right) \right) =\cr
& 6 w + 32 w^{2}  + 110 w^{3}  + 288 w^{4}  + 644 w^{5}  + 1280 w^{6}  + 
  2340 w^{7} + 4000 w^{8} \cr
&  + 6490 w^{9} + 10080 w^{10}   + 15106 w^{11}  + 
  21952 w^{12}   + 31080 w^{13}  \cr
&  + 43008 w^{14}   + 58344 w^{15}   + 
  77760 w^{16} + \ldots  \cr}}
precisely coincides with the one in \local. On the one hand this is very encouraging since 
we managed to count curves of arbitrary genus. On the other hand we were unable so far to extend this counting to, e.g. $(3,n)$ curves. This problem is currently under investigation \nikbarak.

\newsec{Conclusions and speculations.}

\subsec{Higher genus curves}

To extend the techniques of the previous section to cover the higher genera we assume that the
fixed point techniques can be applied there (the obstruction being the non-smoothness of the moduli spaces) (see \enumer\ for discussion). Besides that we need to know 
$\chi (\overline\CM_{g,n})$. 
The virtual Euler characteristics of the uncompactified moduli spaces $\CM_{g,n}$ are 
known \penner\ : for $g >1$
\eqn\euli{\chi_{g,n} = \chi ( \CM_{g,n}) = {{(2- 2g)(2- 2g - 1) \ldots (2 - 2g - (n-1))}\over{ n!}} {{B_{2g}}\over{2g (2g-2)}},}
where $B_{2g}$ is the $2g$-th Bernoulli number.  For genus one:
\eqn\eulio{\chi_{1,n} = {{(-)^{n}}\over{12 n}}, \quad n > 0}
For genus zero:
\eqn\eulize{\chi_{0,n} = {{(-)^{n-1}}\over{n (  n-1)(n-2)}}, \quad n > 2}
This result is however insufficient for our
purposes. We need to know the Euler characteristics of the compactified moduli space with 
labelled punctures. The generating function of the numbers \euli\eulio\eulize\ is:
\eqn\gnfn{\eqalign{ \Sigma( \hbar, \xi ) : = & \sum_{g \geq 0, n \geq 0, 2 - 2g - n < 0} \chi_{g,n} \, {\hbar}^{g} {\xi}^{n} = \cr
= &   {\CS} (\xi) - {{\hbar}\over{12}} {\rm log}(1 + \xi) +  
\sum_{g > 1} {{B_{2g}}\over{2g (2g-2)}} 
 {{\hbar^{g}}\over{(1 + \xi)^{2g-2}}}  \cr}}
Let $\mu = {{(1 + \xi)}\over{\sqrt{\hbar}}} $.
The function $\Sigma (\hbar, \xi) $ has   integral representation: 
\eqn\ps{\Sigma (\hbar, \xi) = - {3\over 4} + 2\hbar^{1\over 2} \mu +  {{\hbar}\over{4}} \left(  2 \mu^2 {\rm log}{\mu} - 3\mu^2    + \int_{\e}^{\mu^{-1}}
 {{ds}\over{s \, {\rm sinh}^2 (s/2) }}   \right),}
where $\e$ is a regulator and one should drop the singular part ${1\over{3}} 
{\rm log} \e  - {2\over{\e^2}}$   in the right hand side.

\medskip
\noindent{\bf The conjecture I.} The Euler characteristics of the compactified moduli spaces of curves $\chi (\overline\CM_{g,n})$ are given by:

\eqn\cnji{\Theta (t,  \hbar) = \sum_{g,n} {1\over{n!}} \chi ( {\overline\CM}_{g,n} ) t^{n} \hbar^{g-1}= 
 {\rm log} \int {{d\xi}\over{\sqrt{2\pi \hbar}}} \exp \,  {1\over{\hbar}}  \left(  \Sigma (\hbar, \xi)  + t \xi \right) }
\medskip

{\bf Remark.} The conjecture is motivated by the following. The expansion of the integral \cnji\ by Feynmann rules generates all possible connected graphs, whose vertices are labelled by the numbers $g_{v}$, s.t. $ g_{v} \geq 0$, and which have the topological characteristics of the stable curve
of genus $g$ with $n$ punctures. The genus $g$ is obtained by summing the internal genera $g_{v}$ and the number of loops in the graph. One should check that the graphs which correspond
to unstable curves $2 - 2g - n \geq 0$ are not generated. In fact, the only possibility is to have a single loop
without vertices, but it has weight zero which follows from the fact that 
$\CS^{\prime\prime}  ( \xi) = -1$ for $\xi = 0$.  In the limit $\hbar \to 0$ we can use saddle point approximation. Then \cnji\ reduces to the statement proven in \manineu.
Notice that the action \ps\ is very much similar to the free energy of Penner model \penner\
although it differs from that  in the unstable terms corresponding to the  genera $0$ and $1$. 
Since we are to take the integral with this action we may say that the Euler characteristics
of the compactified moduli spaces are given by the second quantized Penner model!

\noindent
Now the extension for the stable maps is straightforward: we should sum over  all graphs described in the section $2.2$ and compute:

\eqn\expro{\Theta ( \vec t, \hbar) = \sum_{\Gamma, n, g, \vec d} {{\hbar^{g-1}}\over{\# {\rm Aut}({\Gamma})}} \prod_{v \in V_{\Gamma} } 
\chi ( \overline\CM_{g_{v}, n_{v}} ) \prod_{e \in E_{\Gamma}} e^{- \vec t \cdot \vec d_{e}}}  
By Feynmann rules this is given by  the (path) integral:
\eqn\pathint{\Theta ( \vec t, \hbar)  = {\rm log} \int \prod_{<ij>, d} {\CD} \left( {{\phi_{ij, d}}\over{\sqrt{2\pi \hbar}}} \right) \exp {1\over{\hbar}} S ( \phi, \hbar, \vec t ) }
where the action is the modification of \actn:

\eqn\actnall{\eqalign{S( \phi, \hbar, \vec t ) = & -{1\over 2} \sum_{d, \langle i j\rangle }  \phi_{ij,  d} \phi_{ji, d}  + \cr
+ & \sum_{g, k}  {{\hbar^{g}}\over{k!}} \chi ( \overline\CM_{g,k} )
\sum_{i, \{ ( j_{1}, d_{1}) \ldots (j_{k} , d_{k}) \} } \prod_{\ell = 1}^{k} \phi_{ij_{\ell},  d_{\ell} }
  e^{- {1\over 2} t_{ij_{\ell}} d_{\ell} } \cr}}
Thanks to the conjecture \cnji\ we may rewrite the integral in \pathint\  as the integral over the finite number of auxilliary
variables $\xi_{i}$:
\eqn\auxi{\int \prod_{<ij>, d}  {{{\CD}\phi_{ij, d} \, e^{{{1 + \xi_{i}}\over{\hbar}} \phi_{ij, d} e^{- {1\over 2} t_{ij} d}}}\over{\sqrt{2\pi \hbar}}} \prod_{i}  {{\CD \xi_{i} e^{{\Sigma (\xi_{i} , \hbar) }\over{\hbar}}}\over{\sqrt{2\pi \hbar}}}  e^{-{1\over{2\hbar}} \sum_{d, \langle i j\rangle }  \phi_{ij,  d} \phi_{ji, d} } }
Now the variables $\phi_{ij,d}$ enter the exponent quadratically and can be integrated out:
\eqn\intout{\int \prod_{i}  {{\CD \xi_{i} e^{{\Sigma (\xi_{i} , \hbar) }\over{\hbar}}}\over{\sqrt{2\pi \hbar}}}  e^{{1\over{2\hbar}} \sum_{<ij>} {{(1 + \xi_{i})(1+\xi_{j})}\over{e^{t_{ij}}-1}}} }

Thus we arrive at:

\noindent
{\bf Conjecture II.} The euler characteristics of the moduli spaces of stable holomorphic curves 
in $S$ are given by the logarithm of the integral \intout.

The inclusion of the marked points is as straightforward as in the genus zero case:

\noindent
{\bf Conjecture III.} 
\eqn\mnresall{\eqalign{\qquad  \Theta_{S} (\vec t\vert  t, \hbar) = & \sum_{n, g, \vec d}   \chi (\overline\CM_{g,n, \vec d}) {{t^{n} {\hbar}^{g-1}}\over{n!}} e^{- \vec t \cdot \vec d} =\cr
 {\rm log}    \int \prod_{i}  {{\CD \xi_{i} }\over{\sqrt{2\pi \hbar}}}  & \exp {1\over{2\hbar}} 
\left( \sum_{<ij>} {{(1 + \xi_{i})(1+\xi_{j})}\over{e^{t_{ij}}-1}}  + \sum_{i}  \left( 2 \Sigma( \xi_{i} , \hbar ) + 2 t \xi_{i}  -  t^{2} \right)\right) \cr}}

As a check, in the limit $t_{ij} \to \infty$ in which only degree zero maps survive we get:
$$
\Theta_{S} (\vec t\vert  t, \hbar) \to {\chi (S)} \Theta (t, \hbar) 
$$
as expected.
It is amusing to note that the function $\Sigma(\xi, \hbar)$ also appears in \gopvaf\ in a related context of $M2$-branes, the associated BPS particles and their quantum loops. 

{\bf Remark.} Although it is hard to compute the integral \mnresall\ it is still possible
to get the genus one contributions  without much extra work as long as genus zero
is analysed. We hope to apply this observation elsewhere.

\subsec{The last conlcusions and speculations}

Let us summarize. We studied the genus zero curves which yield the BPS states in the 
compactifications of $M$ -theory on a Calabi-Yau manifold. If the latter contains a surface $S$
then the cuvres in $S$ can be wrapped by $M2$-branes and become BPS particles in five dimensions. The states of the particles correspond to the cohomology of the moduli space of curves. We studied the Euler characteristics of the moduli spaces and found a representation for
the generating function of these numbers in case where $S$ is a toric
variety. 
By making the comparison to the predictions of local mirror symmetry we observed the discrepancy
which has to do with the fact that the moduli space
contains the extra piece - the Jacobians of the
curves\footnote{}{R.~Pandharipande informed me that the topological Euler
characteristics of the moduli space of stable maps 
is not always the right quantity in the sense that it may change under
the variations of the metric on the target space and so on. Instead,
on should work with the Euler classes of the virtual tangent bundle, discussed
 in \rahulp\ and  in a broader physical context in \issues }
Apparently this question is an intriguing and worth further investigations. 
Our method of counting the curves can be applied to the higher genera as
well. 
We presented
our
conjectures concerning this counting.

The last remark concerns the counting of curves in case where the target $S$ is $\IP^1$.
In this case we may find a relation to the proposed string theory of two-dimensional QCD. Namely,  it was suggested in \CMR\ that the latter counts (in the chiral sector) the
euelr characteristics of the moduli space of holomorphic maps of the string worldsheet
into the target space where the large $N$ Yang-Mills theory lives. Unfortunately
it seems that the compactification of the moduli space of maps which is used to reproduce the QCD partition function differs from that of stable maps. 
\bigskip
\bigskip
\centerline{\bf Acknowledgements}
\medskip
I am  grateful to D.~Gaitsgory, B.~Kol, A.~Losev,
Yu.~Manin, R.~Pandharipande, S.~Ramgoolam
and C.~Vafa for  useful discussions and correspondence. 
I am also 
grateful to D.~Gaitsgory for 
his help with communications with P.~Deligne and A.~Vaintrob.  
The research was supported by Harvard Society of Fellows,
partially by NSF under  grant
PHY-9802709, partially by RFFI under grant 98-01-00327 and partially
by grant 96-15-96455 for scientific schools. I am  also grateful to Aspen Center for Physics
for hospitality while part of this work was done. 

\listrefs
\bye